# Effects of cavity-mediated processes on the polarization entanglement of photon pairs emitted from quantum dots

Mukesh Kumar Samal, Divya Mishra and Parvendra Kumar*

Optics and Photonics Centre, Indian Institute of Technology Delhi, Hauz Khas, New Delhi, 110016

*Email: Parvendra@opc.iitd.ac.in

**Abstract**

Semiconductor quantum dots are among the best sources of on-demand entangled photon pairs. The degree of entanglement, however, is generally limited by the fine structure splitting of exciton states. In this paper, we theoretically investigate the generation of polarisation-entangled photon pairs under two-photon excitation and cavity-assisted two-photon emission, both in the weak and strong cavity coupling regimes. We demonstrate and clarify that cavity coupling together with an excitation pulse reduces the degree of entanglement in three different ways. Firstly, in a strong coupling regime, cavity introduces the unequal ac-Stark shift of horizontally and vertically polarised exciton states, which results in the effective splitting of exciton states. Secondly, it induces the cross-coupling between the exciton states even in the weak coupling regime, causing the creation of unfavorable two-photon states. Finally, higher excited states of the cavity modes also contribute to the reduction of entanglement. Therefore, in the setting considered here, cavity coupling, which is generally required for the efficient collection of emitted photons, degrades the entanglement both in weak and strong coupling regimes.

I. **Introduction**

An on-demand and high-brightness source of entangled photon pairs (EPP) constitutes the key ingredient of several protocols in quantum-enabled technologies, including computing, communication, and quantum key distribution [1-6]. Among other quantum emitters such as non-linear crystals, fibres, and atomic and molecular systems, self-assembled quantum dots (QDs) have emerged as a leading platform due to their potential for on-demand generation of maximally EPP with high brightness. Further, their integrability with micro- and photonic crystal cavities makes them suitable candidates for scalable devices. However, generally, in-plane anisotropy of QDs results in the fine structure splitting (FSS) of exciton states, which inhibits the realisation of perfect EPP by introducing the which-path information during the photon generation [7, 8]. Nevertheless, several schemes have been investigated to reduce the exciton splitting, including externally applied strain [9-11]. On the theoretical side, it was predicted that in a near strong coupling regime, cavity-assisted cascaded emission of an initially prepared biexciton state could yield maximally entangled photon pairs even for a finite splitting [12]. Generally, in most experimental settings, a coherent two-photon resonant excitation (TPE) of QD is employed for preparing the biexciton state and simultaneously minimizing the creation of intermediate exciton states [13–15]. Therefore, to optimize the entanglement, it becomes crucial and important to clarify how the simultaneous action of TPE excitation and cavity-assisted emission protocols affect the degree of entanglement. Indeed, very recently, it was shown that the TPE protocol induces an ac-Stark shift of exciton state, resulting the significant degradation of entanglement [16]. However, the effects of TPE were investigated without including the cavity coupling, while in Ref. [12], cavity coupling was considered without including the effect of a two-photon excitation pulse. Note that the coupling of a QD with the appropriate cavity modes is one of the key requirements for the efficient collection of entangled photons.

In this paper, we investigate the generation of polarisation-entangled photon pairs from a QD-cavity system. We employ a state-of-the-art two-photon excitation protocol for biexciton state preparation, and two degenerate cavity modes tuned to half of the biexciton energy for minimising the detrimental effects of FSS [12]. We theoretically demonstrate and quantify that cavity coupling together with a finite TPE pulse degrades the entanglement of emitted photons both in the weak and strong coupling regimes. In a strong coupling regime, cavity introduce a greater ac-Stark shift to a horizontally polarised exciton state compared to that of a vertically polarised exciton state, which results in an effective splitting of exciton states. Naively, it could be assumed that a diagonally polarised pulse would introduce an equal ac Stark shift to both exciton states, resulting in no effective exciton splitting. We, however, show that it introduces a significant cross-coupling between the two exciton states, which quickly degrades the entanglement. It is shown that the emission of more than two photons also occurs, which further contributes to the degradation of entanglement. It is worth mentioning that the coupling of exciton and biexciton states with phonon modes generally affects the degree of polarization entanglement due to the phonon induced processes. In particular authors investigated the effect of the exciton-phonon coupling on the degree of entanglement with respect to the phonon bath temperature and pulse duration [17, 18]. However, recent experiments with InAs QDs in micropillar cavities at low temperatures show that phenomenological dephasing and radiative decay rates fit very well to the experimental results [19, 20]. In this work throughout, we focus on the effect of cavity mediated processes on the polarisation entanglement for a fixed set of pulse duration and experimentally observed decay and dephasing rates at phonon bath temperature 10 K [19].

II. **Theoretical Model**

We model the QD as a four-level system consisting of ground state $|G\rangle$, horizontally $|H\rangle$ and vertically $|V\rangle$ polarized exciton states and a biexciton state $|B\rangle$. The biexciton binding energy, $E_B = 2\hbar(\omega_H - \delta/2) - \hbar\omega_B$, represents the energy difference between two uncorrelated excitons and biexciton state, here $\hbar\omega_H$, $\hbar\omega_B$ and $\delta$ are the energy of horizontally polarized exciton and biexciton states, and FSS of exciton states, respectively [21]. We consider that the exciton and biexciton states are coupled to the two degenerate horizontally (H) and vertically (V) polarized modes of a micropillar cavity. (not shown in Fig. 1).

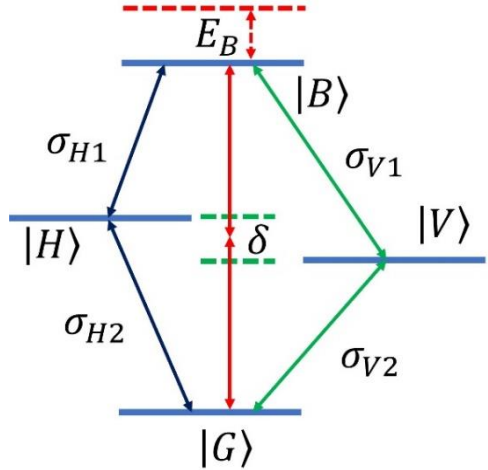

Fig. 1 (Color online) Energy level scheme of a quantum dot, including ground state $|G\rangle$, horizontally, $|H\rangle$, and vertically $|V\rangle$ polarized exciton states and a biexciton state. The transitions between them are represented by the ladder operators $\sigma_{H1} = |H\rangle\langle B|$, $\sigma_{H2} = |G\rangle\langle H|$, $\sigma_{V1} = |V\rangle\langle B|$, and $\sigma_{V2} = |G\rangle\langle V|$. $E_B$ and $\delta$ represent the biexciton binding energy and FSS of the exciton states. Two solid red arrows represent the resonant two-photon excitation of quantum dot.

Furthermore, we assume that the cavity modes and laser pulse (solid red arrows in Fig. 1) frequencies are tuned at half of the biexciton frequency, that is $\omega_c = \omega_l = \omega_B/2$, which satisfy the resonant two-photon interaction. The energy of the ground state $|G\rangle$ is considered to be zero.

The origin of polarization entangled photon pair generation can be understood as follows: The decay of biexciton to the ground state can take place via two equally probable paths, resulting the emission of photons with either horizontal (*H*) or vertical (*V*) polarization. The photons generated via a biexciton-exciton cascade ideally results the formation of a maximally entangled two-photon state, which reads as $|\psi\rangle = (|H_1H_2\rangle + e^{i\phi}|V_1V_2\rangle)/\sqrt{2}$. Note that hereafter, we drop subscript 1 and 2 assuming $|H_1H_2\rangle = |HH\rangle$ and $|V_1V_2\rangle = |VV\rangle$. In the practical conditions, however, the entanglement can be affected by a few uncertainties, including non-zero FSS and emission of more than two photons. The general two-photon density matrix in $|HH\rangle$, $|HV\rangle$, $|VH\rangle$ and $|VV\rangle$ basis reads as [22]:

$$\rho^{tp} = \begin{pmatrix} \alpha_{HH} & \gamma_1 & \gamma_2 & \gamma \\ \gamma_1^* & \beta_{HV} & \gamma_3 & \gamma_4 \\ \gamma_2^* & \gamma_3^* & \beta_{VH} & \gamma_5 \\ \gamma^* & \gamma_4^* & \gamma_5^* & \alpha_{VV} \end{pmatrix}. \qquad (1)$$

The two-photon state $\psi$ corresponds to an ideal case with $|\alpha_{HH}| = |\alpha_{VV}| = |\gamma| = 1/2$ and $\beta_{HV} = \beta_{HV} = \gamma_{1-5} = 0$. For ideal case, the degree of entanglement is usually quantified by the concurrence, $C$, which is defined as $C = 2|\gamma|$; here $\gamma$ is the coherence between two-photon, $|HH\rangle$ and $|VV\rangle$, states. However, in practical situations, the non-vanishing values of $\beta_{HV}$, $\beta_{VH}$, and $\gamma_{1-5}$, result in the degradation of entanglement due to reduced coherence. For such a scenario, the concurrence can be obtained directly from the two-photon density matrix $\rho^{tp}$ by calculating the four eigenvalues $\lambda_j$ of the matrix $M = \rho^{tp}T(\rho^{tp})^*T$, where $(\rho^{tp})^*$ represents the complex conjugated two-photon density matrix, and $T$ is the anti-diagonal matrix with elements $\{-1, 1, 1, -1\}$. The concurrence, $C$ is defined as $C = max\{0, \sqrt{\lambda_1} - \sqrt{\lambda_2} - \sqrt{\lambda_3} - \sqrt{\lambda_3}\}$, where the eigenvalues are sorted in decreasing order, $\lambda_{j+1} \leq \lambda_j$.

III. **Dynamics of QD-Cavity system and Two-photon density matrix elements**

The Hamiltonian of the QD-cavity system in the frame rotating with laser frequency $\omega_l$ and under rotating wave approximation is given as [see appendix]:

$$\widetilde{H}_{RF} = \widetilde{H}_{QD} + \widetilde{H}_{QD-cav} + \widetilde{H}_H + \widetilde{H}_D, \qquad (2)$$

$$\widetilde{H}_{QD} = \frac{\hbar}{2}(E_B/\hbar + \delta)|H\rangle\langle H| + \frac{\hbar}{2}(E_B/\hbar - \delta)|V\rangle\langle V|, \qquad 3(a)$$

$$\widetilde{H}_{QD-cav} = \hbar g\left(a_H^\dagger|G\rangle\langle H| + a_H^\dagger|H\rangle\langle B| + a_V^\dagger|G\rangle\langle V| + a_V^\dagger|V\rangle\langle B| + H.C.\right), \qquad 3(b)$$

$$\widetilde{H}_H = \frac{\hbar\Omega_H(t)}{2}(|G\rangle\langle H| + |H\rangle\langle B| + H.C.), \qquad 3(c)$$

$$\widetilde{H}_D = \frac{\hbar\Omega_D(t)}{2}(|G\rangle\langle H| + |H\rangle\langle B| + |G\rangle\langle V| + |V\rangle\langle B| + H.C.) \qquad 3(d)$$

Here, $\widetilde{H}_{QD}$ is the energy Hamiltonian, $\widetilde{H}_{QD-cav}$, $\widetilde{H}_H$ and $\widetilde{H}_D$ represent the interactions between QD and cavity, QD and horizontally polarized laser pulse, and QD and diagonally polarized laser pulse, respectively. $E_B$ represents the biexciton binding energy, while $a_H^\dagger$ and $a_V^\dagger$ are the photon creation operators into $H$ and $V$ polarized cavity modes, respectively. Moreover, $g$ represents the QD-cavity coupling strength, $\Omega_H(t) = \frac{dE(t)}{\hbar}$ and $\Omega_D(t) = \frac{dE(t)}{\hbar\sqrt{2}}$ are the Rabi frequencies associated with horizontally and diagonally polarized laser pulse, here $d$ is electric dipole moment, which is assumed to be the same for all ground to exciton and exciton to biexciton transitions. The temporal profile of the electric field associated with the laser pulse, $E(t)$ is given as $E(t) = E_0 e^{[-\{(t-t_0)/\tau\}^2]}$. The peak values of Rabi frequencies are defined as: $\Omega_H^p(t = t_0) = \frac{dE_0}{\hbar} = \Omega_H^p$ and $\Omega_D^p(t = t_0) = \frac{dE_0}{\hbar\sqrt{2}} = \Omega_D^p$. $E_0$ is the peak amplitude of the electric field and $\tau$ is related to pulse width at full width at half maxima (FWHM) as $\tau_{FWHM} = 1.177\tau$.

The state of the emitted photons can be obtained from that of the QD-cavity system $\rho$. The time evolution of $\rho$ is computed by solving the following master equation under Born-Markov approximation [23, 24]:

$$\frac{d\rho}{dt} = \frac{i}{\hbar}[\rho, \widetilde{H}_{RF}] + (\mathcal{L}_{QD}^{rr} + \mathcal{L}_{QD}^{d} + \mathcal{L}_{cav})\rho \qquad (4)$$

The first Liouvillian term $\mathcal{L}_{QD}^{rr}$ accounts the radiative relaxation of biexciton and exciton states, which reads as $\mathcal{L}_{QD}^{rr} = \frac{\gamma_B^{rr}}{2}\{\pounds(|H\rangle\langle B|)\rho + \pounds(|V\rangle\langle B|)\rho\} + \frac{\gamma_E^{rr}}{2}\{\pounds(|G\rangle\langle H|)\rho + \pounds(|G\rangle\langle V|)\rho\}$, while $\mathcal{L}_{QD}^{d} = \frac{\gamma_B^d}{2}\{\pounds(|B\rangle\langle B|)\rho\} + \frac{\gamma_E^d}{2}\{\pounds(H\rangle\langle H|)\rho + \pounds(|V\rangle\langle V|)\rho\}$ incorporates the dephasing of biexciton and exciton states, and $\mathcal{L}_{cav} = \frac{\kappa}{2}\{\pounds(a_H)\rho + \pounds(a_V)\rho\}$ describes the cavity decay rates. Here, $\pounds(A)\rho = 2A\rho A^\dagger - A^\dagger A\rho - \rho A^\dagger A$. The terms $\gamma_B^{rr}$ and $\gamma_E^{rr}$ are the radiative decay rates of biexciton and exciton ($H$ and $V$ polarized) states, respectively, while $\gamma_B^d$ and $\gamma_E^d$ are the dephasing rates of biexciton and exciton ($H$ and $V$ polarized) states, respectively. The radiative decay rates usually depend on the local density of states, while dephasing depends on the exciton-phonon coupling [25, 26]. $\kappa$ represents the decay rate of $H$ and $V$ polarized cavity photons.

The elements of two-photon density matrix, which is usually reconstructed in experiments by means of tomographic method, theoretically correspond to a specific set of time-averaged second-order correlation functions. For calculating the concurrence, we computed all such second-order correlation functions by numerically solving Eq. 4 and by employing the quantum regression theorem. For example, a diagonal two-photon density matrix element, $\alpha_{HH}$ reads as $\alpha_{HH} = B\int_0^\infty dt \int_0^\infty dt' \langle a_H^\dagger(t) a_H^\dagger(t') a_H(t') a_H(t)\rangle$, where $B$ is a normalization constant such as $B(\alpha_{HH} + \beta_{HV} + \beta_{VH} + \alpha_{VV}) = 1$. The off-diagonal terms in $\rho^{tp}$ are calculated as follows: $\langle \mu\, \nu|\rho^{tp}|\xi\, \zeta\rangle = B\int_0^\infty dt \int_0^\infty dt' \langle a_\mu^\dagger(t) a_\nu^\dagger(t') a_\zeta(t') a_\xi(t)\rangle$, where $\mu, \nu, \xi, \zeta \in H, V$, specifically, coherence $\gamma$ is computed as $\gamma = \langle H\,H|\rho^{tp}|V\,V\rangle = B\int_0^\infty dt \int_0^\infty dt' \langle a_H^\dagger(t) a_H^\dagger(t') a_V(t') a_V(t)\rangle$. We use the typical values of InAs QDs parameters as:, $\gamma_E^{rr} = 1\ \mu eV$, $\gamma_B^{rr} = 2\ \mu eV$, $\gamma_E^d = 2\ \mu eV$, $\gamma_B^d = 4\ \mu eV$, $E_B = 1\ meV$ [19, 20, 27, 28]. The other parameters employed throughout this paper are $\tau_{FWHM} = 3.53\ ps$ and $\kappa = 65\ \mu eV$. Note that we chose a fixed value of cavity decay rate $\kappa$ throughout this work and different cavity coupling strength $g$ for investigating the degree of entanglement in weak $(g < \kappa)$ and strong $(g > \kappa)$ regimes. The optimized peak Rabi frequencies for creating the maximally populated biexciton state are $\Omega_H^p = \Omega_D^p = 1\ meV$. The values of other parameters are mentioned appropriately in the text.

IV. **Results and Discussions**

The fine structure splitting of exciton states is unique to every single QD and depends on its shape and strain environment [29]. Generally, FSS in InAs quantum dots lies in the range of tens of $\mu eV$ [29-31]. In Fig. 2 (a), we show the evolution of concurrence as a function of $g$ for an initially prepared biexciton state, i.e., $\rho_{BB} = 1$ at three different values of $\delta = 0$ (solid red line), $\delta = 20$ (dashed blue line) and $40\ \mu eV$ (dashed-dotted green line). It can be observed that more than 95 % concurrence can be obtained even with a finite value of FSS, $\delta = 40\ \mu eV$ in strong coupling regime, $g > \kappa$. Next, in Fig. 2 (b), we consider a practical situation by assuming the ground state as the initial state of QD, i.e., $\rho_{GG} = 1$. Furthermore, we include the coupling of a horizontally polarized laser pulse for preparing the QD into the biexciton state under two-photon resonant excitation. It can be observed that under these realistic conditions, the value of concurrence is reduced significantly at the larger values of $g$.

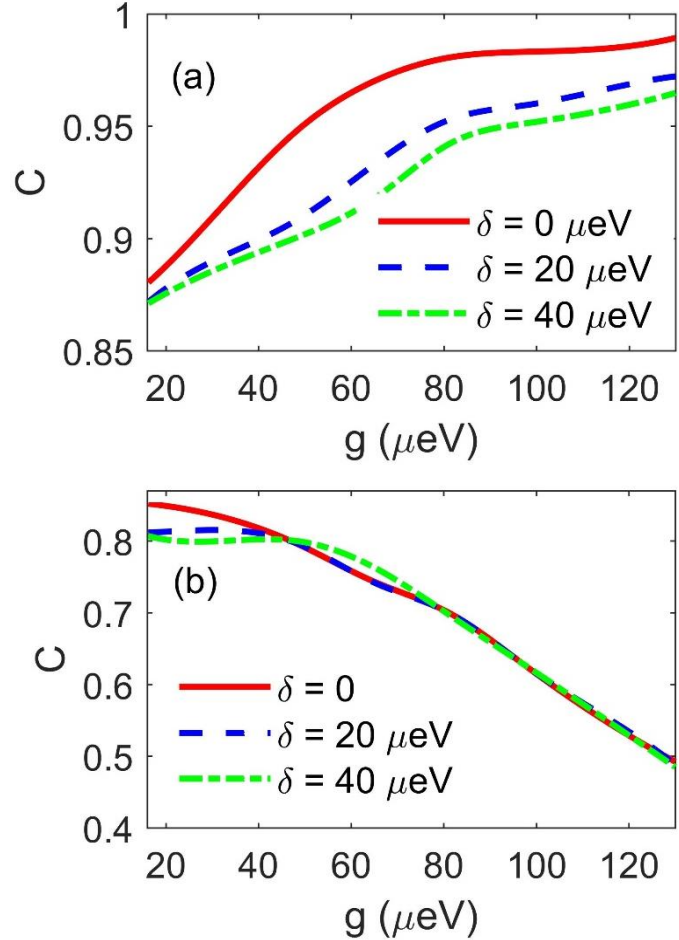

Fig. 2 (Color online) Evolution of concurrence as a function of $g$ (a) $\delta = 0$ (solid red line), $\delta = 20$ (dashed blue line) and $\delta = 40\ \mu eV$ (dashed-dotted green line) with $\rho_{BB} = 1$, (b) $\delta = 0$ (solid red line), $\delta = 20$ (dashed blue line) and $\delta = 40\ \mu eV$ (dashed-dotted green line) with $\rho_{GG} = 1$.

It is clear from Fig. 2(b) that the cavity coupling together with an excitation pulse significantly reduces the concurrence. We find that the degradation of entanglement occurs due to the reduced coherence between $|HH\rangle$ and $|VV\rangle$ states. We identified and illustrated that the cavity-mediated process, including, cavity-induced ac-Stark shift of exciton states, cross-coupling between $H$ and $V$ polarised exciton states, and emission of more than two photons are the main cause of reduced coherence.

In Fig. 3, we plot the cavity induced ac-Stark shifts $\Delta_{HH} = 2\langle a_H^\dagger a_H\rangle \frac{g^2}{\delta_H}$ and $\Delta_{VV} = 2\langle a_V^\dagger a_V\rangle \frac{g^2}{\delta_V}$ of $H$ and $V$ polarised exciton states, respectively [32, 33]. Here, $\langle a_H^\dagger a_H\rangle$ and $\langle a_V^\dagger a_V\rangle$ are the average photon number in $H$-polarized and $V$-polarized cavity modes, respectively, $\delta_H = (E_B/\hbar + \delta)/2$ and $\delta_V = (E_B/\hbar - \delta)/2$. In Fig. 3(a), we show the evolution of $\Delta_{HH}$ and $\Delta_{VV}$ at $\delta = 0$ for an initially prepared biexciton state, i.e., $\rho_{BB} = 1$. It can be observed that both $\Delta_{HH}$ and $\Delta_{VV}$ are equal for the entire range of chosen values of $g$, hence cavity-induced effective splitting of the exciton states does not appear. Consequently, the value of concurrence increases as a function of $g$ [see Fig. 2(a)]. However, as we show in Fig. 3(b), the Stark shift of $H$-polarized exciton state dominates, especially for $g > 65\ \mu eV$ under TPE of QD.

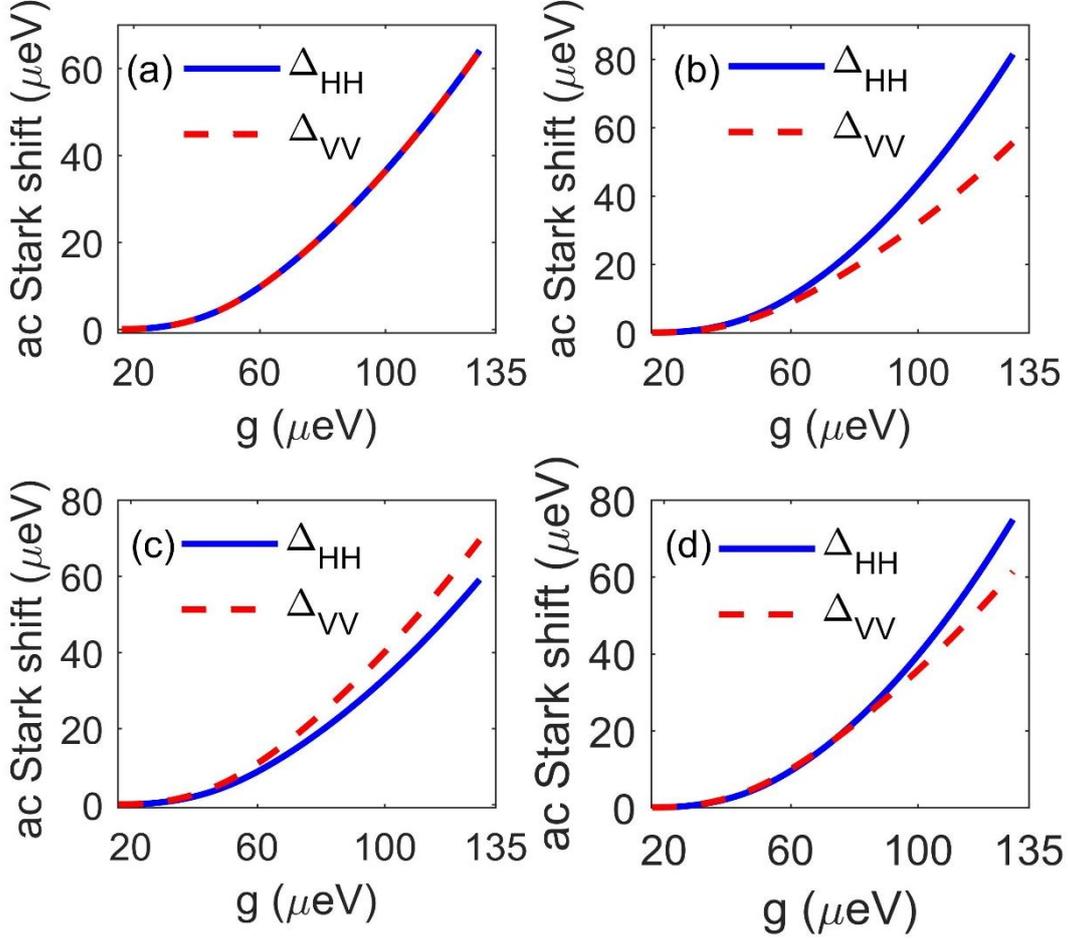

Fig. 3 (Color online) Evolution of the cavity induced Stark shifts $\Delta_{HH}$ and $\Delta_{VV}$ as a function of $g$. (a) $\delta = 0$ and $\rho_{BB} = 1$ without TPE (b) $\delta = 0$, $\rho_{GG} = 1$ (c) $\delta = 40\ \mu eV$, $\rho_{BB} = 1$ and (d) $\delta = 40\ \mu eV$ and $\rho_{GG} = 1$ with a $H$-polarized laser pulse induced TPE.

This simply occurs because $H$-polarized laser pulse only interacts with $|H\rangle$ state, thereby assisting the emission of more photons into the $H$-polarized cavity mode, resulting the greater value of $\langle a_H^\dagger a_H \rangle$ compared to that of $\langle a_V^\dagger a_V \rangle$. Consequently, the unequal Stark shifts of $H$ and V-polarized exciton states result in an effective splitting of the exciton states, causing the degradation of entanglement. Moreover, it can be observed from Fig. 3(c) that there exists a cavity-mediated Stark shift even for an initially prepared biexciton state i.e., $\rho_{BB} = 1$ at $\delta = 40\ \mu eV$. Note that here, $\Delta_{VV} > \Delta_{HH}$, just because of $\delta_V < \delta_H$. Next in Fig. 3(d), the Stark shift follows the same trend as shown in Fig. 3(b). Overall, from Fig. 3, it is clear that cavity-mediated Stark shift under TPE of QD results in the effective splitting of exciton states, particularly in the strong coupling regime ($g > \kappa$), which in turn adds to the which-path information. In the two-photon density matrix picture, this manifests itself in the form of reduced coherence between $|HH\rangle$ and $|VV\rangle$ states, eventually causing the reduction of concurrence.

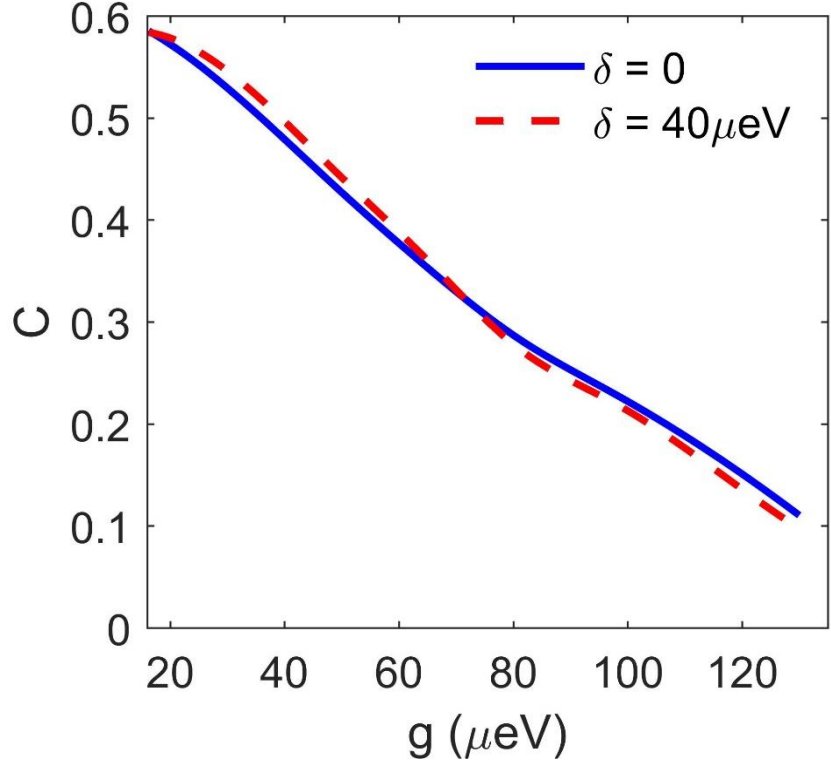

Fig. 4 (Color online) Evolution of concurrence as a function of $g$ under two-photon excitation using a diagonally polarized laser pulse.

Following the analysis of Figs. 3(b) and (d), one can naively assume that the greater Stark shift of $H$-polarized exciton state can be circumvented by simultaneously employing a $V$-polarized laser pulse. To test this, we employ a diagonally polarized laser pulse, i.e., a laser with equal $H$ and $V$ polarized electric field components. Indeed, we find that the Stark shifts $\Delta_{HH}$ and $\Delta_{VV}$ are exactly equal for $\delta = 0$ (not shown).

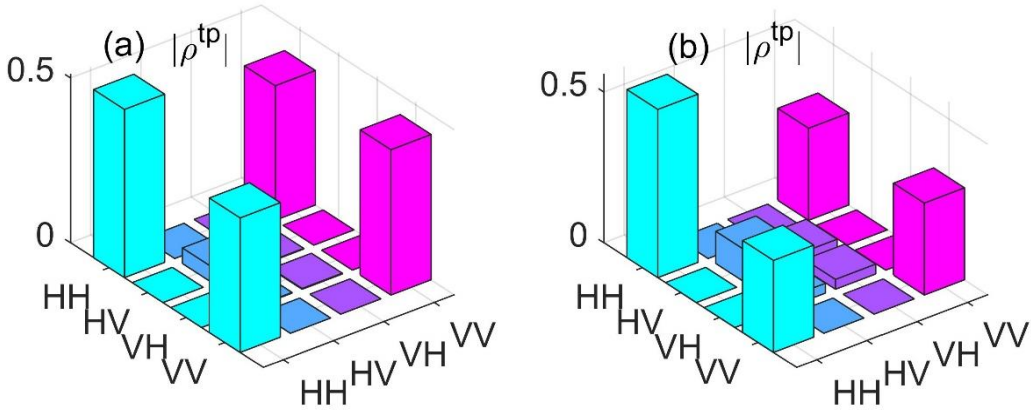

Fig. 5 (Color online) Two-photon density matrices for a $H$-polarized laser pulse induced two-photon excitation (a) $\delta = 0, g = 45\ \mu eV$ and (b) $\delta = 0, g = 130\ \mu eV$.

Nevertheless, we observe from Fig. 4 that the concurrence drops significantly for both $\delta = 0$ and $40\ \mu eV$. Therefore, the significant reduction in concurrence cannot be attributed to the cavity-mediated Stark shift. We find that this is related to the cross-coupling between the exciton states and emission of more than two photons. This is illustrated and clarified in Figs.

7 and 9, respectively. In Fig. 5, we show the two-photon density matrix for a $H$-polarized laser pulse-induced two-photon excitation. The creation of a two-photon state $|HV\rangle$ can be observed for $g = 45$ and $130\ \mu eV$. Additionally, a two-photon state $|VH\rangle$ is also created at $130\ \mu eV$. It can be observed from Fig. 5 that the coherence between $|HH\rangle$ and $|VV\rangle$ gets reduced partly due to the creation of two-photon states with orthogonal polarization, which also contributes to the decrement of concurrence.

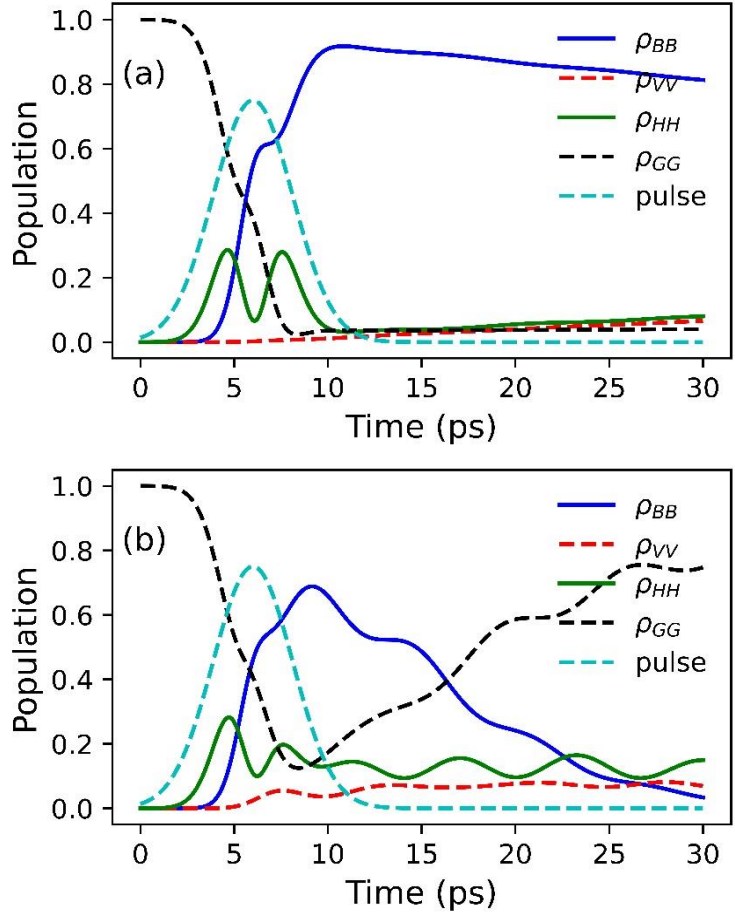

Fig. 6 (Color online) Temporal profile of laser pulse (cyan dashed line), ground (black dashed line), $H$-polarized exciton (green solid line), $V$-polarized exciton (red dashed line), and biexciton (solid blue line) states population at (a) $\delta = 0, g = 45\ \mu eV$ and (b) $\delta = 0, g = 130\ \mu eV$.

For further clarifying the formation of $|HV\rangle$ and $|VH\rangle$ states, we show the temporal dynamics of ground, $H$- and $V$-polarized exciton, and biexciton states population denoted by $\rho_{GG}$, $\rho_{HH}$, $\rho_{VV}$ and $\rho_{BB}$, respectively. It can be observed from Fig. 6(a) that the excitation and deexcitation of $H$-polarized exciton state occurs during the laser pulse itself.

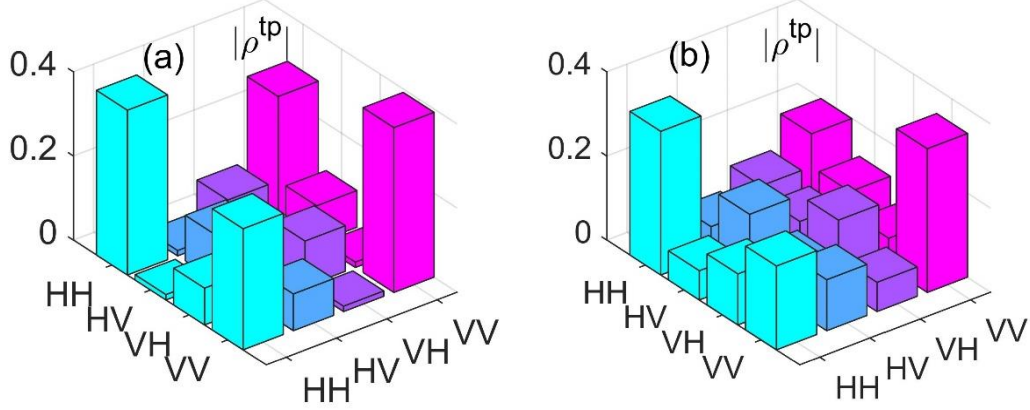

Fig. 7 (Color online) Two-photon density matrices for a diagonally polarized laser pulse induced two-photon excitation at (a) $\delta = 0, g = 45\ \mu eV$ and (b) $\delta = 0, g = 130\ \mu eV$.

The formation of $|HV\rangle$ state can be understood as follows. The decay of biexciton state creates an $H$-polarized photon in the cavity mode, while the simultaneous decay of $V$-polarized exciton state creates a $V$-polarized photon into the cavity mode, resulting a two-photon state with orthogonal polarization. Similarly, in Fig. 6 (b), the oscillations of $\rho_{HH}$ and $\rho_{VV}$ persist even after the laser pulse, causing the formation of both $|HV\rangle$ and $|VH\rangle$ states. It can also be observed from Fig. 6 that compared to $g = 45\ \mu eV$, the value of $\rho_{GG}$ is greater ($\rho_{BB}$ is lesser) for $g = 130\ \mu eV$. This is due to the greater coupling strength of cavity mode with the QD exciton and biexciton states, consequently these states quickly and predominately decay by emitting the photons into the cavity modes. Moreover in Fig. 6 (b), the sustained oscillations of $\rho_{HH}$ and $\rho_{VV}$ occurs because of the reabsorption of cavity photons by QD at $g = 130\ \mu eV$. Next, in Fig. 7, we show the two-photon density matrices under TPE with a diagonally polarized laser pulse at two different values of $g$. It can be observed from Fig. 7(a) and 7(b) that all matrix elements take non-zero values, especially at $g = 130\ \mu eV$.

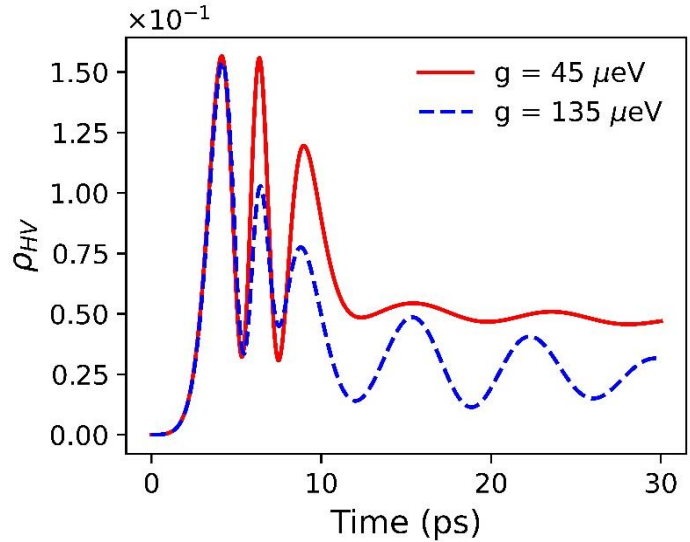

Fig. 8 (Color online) Evolution of coherence between $H$- and $V$-polarized exciton states $\rho_{HV}$ at $g = 45\ \mu eV$ (solid red line) and $g = 130\ \mu eV$ (blue dashed line).

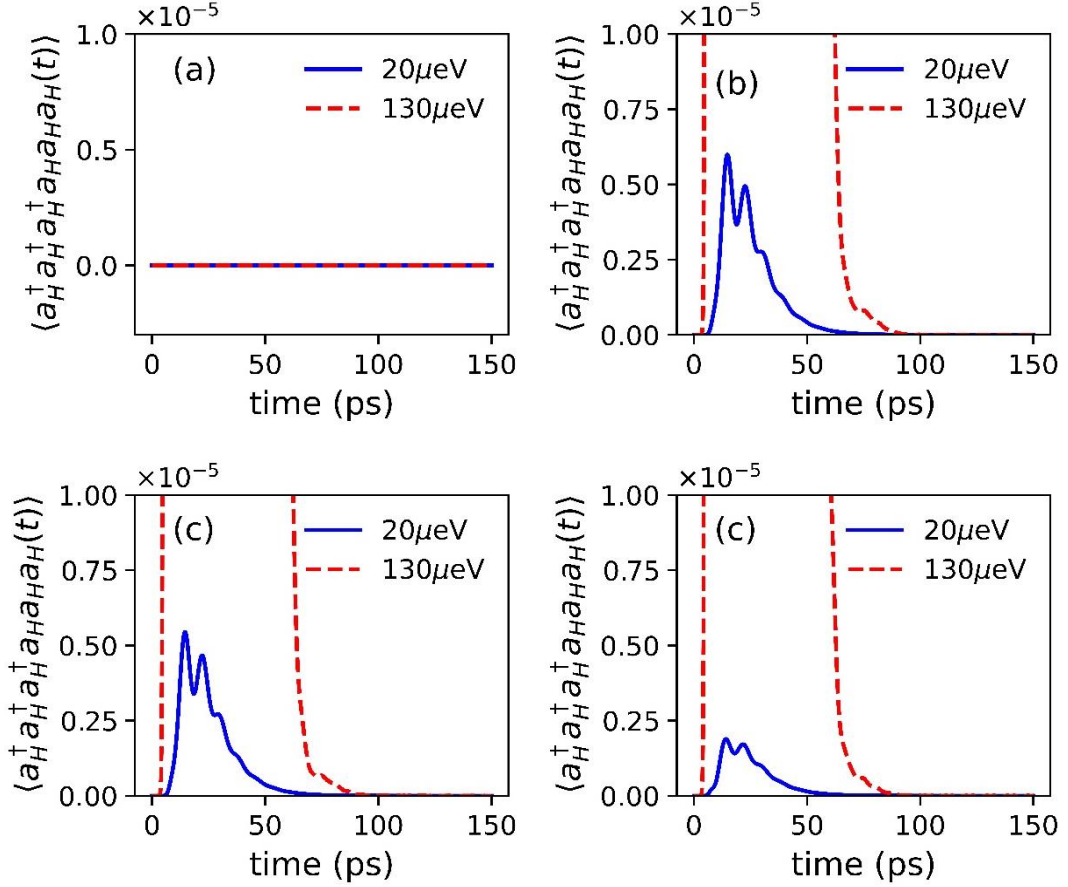

Fig. 9 (Color online) Evolution of equal-time third order correlation function (a) $\delta = 0$, $\rho_{BB} = 1$ (b) $\delta = 0$, $\rho_{GG} = 1$, (c) $\delta = 40\ \mu eV$, $\rho_{GG} = 1$, for a H polarized laser pulse, (d) ) $\delta = 0$, $\rho_{GG} = 1$ for a diagonally polarized laser pulse.

In particular, significant populations in $|HV\rangle$ and $|VH\rangle$ states occur due to the cross-coupling between the exciton states as evident from the non-zero value of coherence $\rho_{HV}$, shown in Fig. 8. The finite population of $|HV\rangle$, $|VH\rangle$ and other off-diagonal two-photon states amounts to the significant reduction in concurrence as can be seen from Fig. 4. It may be noted that these results are in sharp contrast to the results reported in Ref. [16], wherein the concurrence remains similar for horizontal and diagonal polarized laser pulses induced two-photon excitation. This is precisely due to the inclusion of the finite cavity coupling in this work. In Fig. 8, we show the evolution of coherence $\rho_{HV}$ at two different values of $g$. It can be observed that the significant oscillation of coherence persists for a longer times, especially at $g = 130\ \mu eV$. The finite value of $\rho_{HV}$ reflects the cross-coupling between H- and V- polarized exciton states, which causes the flipping of the polarization states of the biexciton emitted photons. Since, the oscillation of coherence persists for a longer time at $g = 130\ \mu eV$ that's why there is a greater value population of both $|HV\rangle$ and $|VH\rangle$ states compared at $g = 45\ \mu eV$ shown in Fig. 7(a). Finally, in Fig. 9, we show the evolution of equal-time third order correlation function (ETTOCF) $\langle a_H^\dagger a_H^\dagger a_H^\dagger a_H a_H a_H(t)\rangle$, which takes nonzero values whenever there are more than two horizontally polarized photons in the cavity modes. If there are two or less than two photons in the cavity modes, then its value is zero. Therefore, $\langle a_H^\dagger a_H^\dagger a_H^\dagger a_H a_H a_H(t)\rangle$ represents the temporal evolution of more than two-photon population in the cavity mode. It could be observed from Fig. 9 (a) that the ETTOCF is always zero for both values of $g$. However, in the presence of excitation laser pulse ETTOCF takes the non-zero values for both $g = 20$ and $130 \, \mu eV$. It can be understood that this occurs due to the cavity coupling mediated re-absorption of cavity photons by QD and subsequent emission into the cavity modes. The emission of more than two photons results in the reduction of coherence between $|HH\rangle$ and $|VV\rangle$ states and also contributes to the degradation of entanglement.

## V. Conclusion

To conclude, we have shown that under two-photon excitation of quantum dot, cavity-mediated processes hinder the realization of maximum entanglement particularly in strong coupling regime by reducing the coherence in three different ways. Firstly, it is shown that the achievable entanglement gets reduced due to the cavity-induced ac Stark shift-driven effective splitting of exciton states. Secondly, we have demonstrated that the cavity-mediated creation of a two-photon state with orthogonal polarisation and the emission of more than two photons also reduce the degree of entanglement. Finally, we have shown that the two-photon excitation with a diagonally polarised laser pulse introduces a significant cross-coupling between the two exciton states, which quickly reduces the degree of entanglement. Specifically, we have shown that the cavity-mediated Stark shift is almost zero for weak cavity coupling, but the creation of more than two-photon states and cross coupling between exciton states exist for both weak and strong coupling regimes. However, in a weak coupling regime, their magnitudes are smaller, resulting in less reduction of entanglement. Finally, note that in this work, we focused on the effects of cavity-mediated processes for fixed values of decay and dephasing rates. The significance of phonon-bath-induced incoherent processes, including decay and dephasing rates, at different temperatures will be investigated separately.

**Acknowledgement**
D.M. and M.K.S. gratefully acknowledges support of a research fellowship from IIT Delhi and JRF(NET)-UGC, government of India. P.K. acknowledges support through new faculty seed grant IIT Delhi.**APPENDIX: Derivation of the rotating frame Hamiltonian [Eq. 2]**

The Hamiltonian of the QD-cavity system shown in Fig. 1 can be written as follows:

$$\boldsymbol{H} = \boldsymbol{H_{QD}} + H_{cav} + H_{QD-cav} + H_H + H_D, \tag{A1}$$

$$H_{QD} = \hbar\omega_H|H\rangle\langle H| + \hbar\omega_V|V\rangle\langle V| + \hbar\omega_B|B\rangle\langle B|, \tag{A2}$$

$$H_{cav} = \omega_c a_H^\dagger a_H + \omega_c a_V^\dagger a_V, \tag{A3}$$

$$H_{QD-cav} = \hbar g(|G\rangle\langle H| + |H\rangle\langle B|+|G\rangle\langle V| + |V\rangle\langle B| + H.C.)\big(a_H^\dagger + a_V^\dagger + H.C.\big), \tag{A4}$$

$$H_H = -\frac{dE(t)}{2}(|G\rangle\langle H| + |H\rangle\langle B| + H.C.)\big(e^{-i\omega_l t} + H.C.\big), \tag{A5}$$

$$H_D = -\frac{dE(t)}{2\sqrt{2}}(|G\rangle\langle H| + |H\rangle\langle B|+|G\rangle\langle V| + |V\rangle\langle B| + H.C.)\big(e^{-i\omega_l t} + H.C.\big), \tag{A6}$$

here, $\omega_H$, $\omega_V$ and $\omega_B$ are the frequencies of horizontally, vertically polarized exciton and biexciton states, respectively. $\omega_c$ is the frequency of horizontally and vertically polarized degenerate cavity modes, $g$ represents the coupling strength of the cavity modes with the exciton and biexciton states and $d$ is the electric dipole moment as stated in the main text. Hamiltonian in a frame rotating with frequency $\omega_l$ is calculated as follows:

$$\widetilde{H}_{RF} = \widehat{U}H\widehat{U}^\dagger + i\widehat{U}^\dagger \frac{d\widehat{U}}{dt}, \tag{A7}$$

where unitary operator reads as $\widehat{U} = e^{i\omega_l t\left(|H\rangle\langle H|+|V\rangle\langle V|+2B\rangle\langle B|+a_H^\dagger a_H+a_V^\dagger a_V\right)}$. We use the Baker–Hausdorf lemma: $e^{i\alpha \widehat{A}}\widehat{B}e^{-i\alpha \widehat{A}} = \widehat{B} + i\alpha[\widehat{A},\widehat{B}] + \frac{(i\alpha)^2}{2!}[\widehat{A},[\widehat{A},\widehat{B}]] + \cdots$ and rotating wave approximation [34] for deriving the $\widetilde{H}_{RF}$, which is given as

$$\widetilde{H}_{RF} = \widetilde{H}_{QD} + \widetilde{H}_{cav} + \widetilde{H}_{QD-cav} + \widetilde{H}_H + \widetilde{H}_D, \tag{A8}$$

$$\widetilde{H}_{QD} = \hbar(\omega_H - \omega_l)|H\rangle\langle H| + \hbar(\omega_V - \omega_l)|V\rangle\langle V| + \hbar(\omega_B - 2\omega_l)|B\rangle\langle B|, \tag{A9}$$

$$H_{cav} = \hbar(\omega_c - \omega_l)a_H^\dagger a_H + \hbar(\omega_c - \omega_l)a_V^\dagger a_V, \tag{A10}$$

$$\widetilde{H}_{QD-cav} = \hbar g\left(a_H^\dagger|G\rangle\langle H| + a_H^\dagger|H\rangle\langle B| + a_V^\dagger|G\rangle\langle V| + a_V^\dagger|V\rangle\langle B| + H.C.\right), \tag{A11}$$

$$\widetilde{H}_H = \frac{\hbar\Omega(t)}{2}(|G\rangle\langle H| + |H\rangle\langle B| + H.C.), \tag{A12}$$

$$\widetilde{H}_D = \frac{\hbar\Omega(t)}{2\sqrt{2}}(|G\rangle\langle H| + |H\rangle\langle B| + |G\rangle\langle V| + |V\rangle\langle B| + H.C.) \tag{A13}$$

The Eq. A8 gets convert to Eq. 2 of the main text after putting $\omega_c = \omega_l$ in Eq. A10 and $\omega_l = \omega_B/2$, $\omega_B = 2(\omega_H - \delta/2) - E_B/2$ in Eq. A9.